\input harvmac 
\input pictex
\overfullrule=0pt
\def\half{{\textstyle{1\over 2}}}

\def\neqno#1{\eqnn{#1} \eqno #1}

\Title{hep-th/0103103, UTPT-01-06} {\vbox{
\centerline{The Randall-Sundrum Scenario with} \vskip8pt
\centerline{an Extra Warped Dimension} }}
\centerline{Hael Collins\footnote{$^\dagger$}{{\tt 
hael@physics.utoronto.ca}} and Bob 
Holdom\footnote{$^\ddagger$}{{\tt bob.holdom@utoronto.ca}} }
\medskip
\vbox{\it \centerline{Department of Physics} \vskip-2pt
\centerline{University of Toronto} \vskip-2pt
\centerline{Toronto, Ontario M5S 1A7, Canada}}

\bigskip
\bigskip
\centerline{Abstract}

\medskip 
{\baselineskip=10pt\centerline{\vbox{\hsize=.8\hsize\ninepoint\noindent  
We investigate a scenario with two four-branes embedded in six dimensions. 
When the metric is periodic and compact in one of the dimensions parallel to
the branes, the value of the effective cosmological constant for the remaining
five dimensions can assume a variety of values, determined by the dependence
of the metric on the sixth dimension.  The picture that emerges resembles the
Randall-Sundrum model but with an extra warped dimension that allows the usual
brane-bulk fine tuning to be satisfied {\it without\/} finely tuning any of
the parameters in the underlying six dimensional theory.  Although the action
contains terms with four derivatives of the metric, we show that when the
branes have a finite, natural thickness, such terms have only a small effect
on the Randall-Sundrum structure.  The presence of these four derivative terms 
also allows a configuration that resembles that produced by a domain wall but
which results from gravity alone.  }}}

\Date{March, 2001}
\baselineskip=12pt

\newsec{Introduction.}

Two of the most enigmatic features of the universe are the weakness of gravity
compared to electroweak interactions---the hierarchy problem---and the small
size of the cosmological constant.  A recent approach 
\ref\rsa{L.~Randall and R.~Sundrum, ``A large mass hierarchy from a small
extra dimension,'' Phys.\ Rev.\ Lett.\ {\bf 83}, 3370 (1999)
[hep-ph/9905221].} 
to the hierarchy problem has observed that the presence of a small warped
extra dimension could naturally produce an exponential hierarchy between the
scales of gravitational and Standard Model interactions.  The new element in
this and related 
\ref\mmdim{N.~Arkani-Hamed, S.~Dimopoulos and G.~Dvali, ``The hierarchy
problem and new dimensions at a millimeter,'' Phys.\ Lett.\ B {\bf 429}, 263
(1998) [hep-ph/9803315] and 
I.~Antoniadis, N.~Arkani-Hamed, S.~Dimopoulos and G.~Dvali, ``New dimensions
at a millimeter to a Fermi and superstrings at a TeV,'' Phys.\ Lett.\ B {\bf
436}, 257 (1998) [hep-ph/9804398].} 
scenarios is the introduction of solitonic three-dimensional hypersurfaces, or
$3$-branes, to which the Standard Model fields are confined while gravity
propagates in all the dimensions.  These models still require a fine-tuning to
produce a low energy effective $3+1$ dimensional theory with no cosmological
constant.

Extra dimensions might also provide a framework for addressing the
cosmological constant problem.  Instead of setting the cosmological constant
to an unnaturally small value, we can demand only that the theory should admit
a nearly flat effectively $3+1$ dimensional theory below some high energy
scale---regardless of the value of the cosmological constant.  This picture
was introduced by Rubakov and Shaposhnikov with a six dimensional model 
\ref\rs{V.~A.~Rubakov and M.~E.~Shaposhnikov, ``Extra Space-Time Dimensions:
Towards A Solution To The Cosmological Constant Problem,'' Phys.\ Lett.\ B
{\bf 125}, 139 (1983).}.  
The idea is that the cosmological constant distorts, or warps, some or all of
the extra dimensions while leaving the theory with a $3+1$ dimensional
Poincar\' e symmetry.  If this idea is extended so that this warping is
accomplished with a metric that is both smooth and periodic in the extra
dimensions, then there is no need to cut off the space or to encounter
singularities in the extra dimensions.

An explicit realization occurs in $4+1$ dimensions 
\ref\warped{H.~Collins and B.~Holdom, ``The cosmological constant and warped
extra dimensions,'' hep-th/0009127.} 
when the metric is smooth, non-singular and periodic in the extra dimension. 
We can then choose the extra dimension to be compact with its size given by
the period.  At large distances compared to this period, the universe appears
four-dimensional.  The $4d$ cosmological constant is determined by both the
$5d$ cosmological constant and the geometry of the extra dimension. 
Therefore, we can achieve $3+1$ dimensional Poincar\' e invariance even when
the $5d$ cosmological constant is not zero by choosing the solution to the
field equations with the appropriate behavior in the extra dimension. 
However, some further mechanism is still required to explain why this
particular solution should be preferred.

This note intends to combine these ideas into a scenario that incorporates the
Randall-Sundrum picture $\rsa$ but without finely tuning any of the parameters
in the action.  The scenario starts with an effective action for gravity in
six dimensions including terms with up to four derivatives of the metric.  It
also includes two parallel $4$-branes which are compact in one dimension and
extend infinitely in the other three.  With some mild bounds on the parameters
in the action, we find that the equations of motion allow the geometry of the
resulting universe to contain a five-dimensional anti-de Sitter (AdS$_5$)
subspace with a warped metric that is periodic in the sixth dimension.  

A generic set of four derivative terms in the action apparently implies an
infinite tension on the branes; however, this singularity appears as an
artifact of the vanishing thickness of the branes.  When the brane has a
finite thickness, it is possible to show explicitly that the higher derivative
terms can be neglected.  An action that contains {\it only\/}
gravity---including these four derivative terms but without any scalar
fields---also admits solutions in which gravity is localized about a
codimension one hypersurface.  Far from this hypersurface, the metric
approaches an AdS metric as in the second Randall-Sundrum model 
\ref\rsb{L.~Randall and R.~Sundrum, ``An alternative to compactification,''
Phys.\ Rev.\ Lett.\ {\bf 83}, 4690 (1999) [hep-th/9906064].}; 
however, it is the four derivative terms and not a brane or a scalar field
that effects this localization of gravity.

\newsec{Preliminaries and effective action descriptions.}

Randall and Sundrum $\rsa$ proposed that if the universe were to consist of
two $3$-branes bounding a bulk region of five-dimensional anti-de Sitter
space-time, then the redshift induced by the bulk metric at one of the branes
could generate an exponential hierarchy between the Planck scale and the scale
of electroweak symmetry breaking.  The action for their scenario contains an
Einstein-Hilbert term for the bulk,
$$S_{\rm bulk}^{\rm RS} = M_5^3 \int d^4xdr\, \sqrt{-\hat g} \left[
2\Lambda_{\rm RS} + R \right] , \neqno\rsaction$$
while the branes located at $r=0$ and $r=r_c$ only contribute through their
surface tensions,
$$S_{\rm branes}^{\rm RS} = M_5^3 \int_{r=0} d^4x\, \sqrt{-\hat h} \left[ - 2
\sigma_{\rm RS} \right] + M_5^3 \int_{r=r_c} d^4x\, \sqrt{-\hat h} \left[ 2
\sigma_{\rm RS} \right] . \neqno\rssurf$$
Here $\hat g_{MN}$ is the metric for AdS$_5$,
$$\hat g_{MN}\, dx^Mdx^N = e^{-2|r|/\ell} \eta_{\mu\nu}\, dx^\mu dx^\nu + dr^2
, \neqno\rsmetric$$
and $\hat h_{\mu\nu}$ is the induced metric on the branes.  We shall denote
the usual space-time directions by $x^\mu$, with $\mu,\nu\, \ldots = 0,1,2,3$, 
and $r$ describes the direction orthogonal to the branes with $M,N\, \ldots =
0,1,2,3,r$.  $M_5$ denotes the bulk Planck mass.  The bulk Einstein equations
determine $\Lambda_{\rm RS} = {6\over\ell^2}$ and the specific choice of
$\sigma_{\rm RS} = {6\over\ell}$ for the brane tensions is necessary for the
low energy four-dimensional theory to be free of a cosmological constant.

As the cosmological constant and the surface tension appear in the action,
they represent fundamental parameters of the theory and we have no reason {\it
a priori\/} that the fine-tuning condition is satisfied.  If instead the
quantities that appear in the action arise from some more fundamental theory,
then it might be possible for a dynamical mechanism to exist that favors
solutions in which the low energy, four-dimensional theory is nearly flat.

We can adapt the picture developed in $\warped$ without branes to one which
resembles the Randall-Sundrum construction but where the AdS$_5$ length is not
uniquely determined by the higher-dimensional cosmological constant.  The
structure for such a model would include {\it two\/} extra dimensions---one
small periodic dimension to avoid fine-tuning the cosmological constant and a
second to generate the electroweak-Planck hierarchy.
As in $\warped$, we consider gravity as an effective theory, expanded in
powers of derivatives, with a scalar field $\phi$,
$$\eqalign{S_{\rm bulk} 
&= M_6^4 \int d^4xdrdy\, \sqrt{-g} \left( 2\Lambda + R + a R^2 + b
R_{ab}R^{ab} + c R_{abcd}R^{abcd} + \cdots \right) \cr
&\quad
+ M_6^4 \int d^4xdrdy\, \sqrt{-g} \left( - \half \nabla_a\phi \nabla^a\phi +
\Delta{\cal L} \right) . \cr}  \neqno\action$$
Here $y$ represents the coordinate of the extra periodic dimension, with
$a,b,\ldots = 0,1,2,3,r,y$.  $\Lambda$ and $M_6$ denote the cosmological
constant and the six dimensional Planck mass respectively.  
$$\beginpicture
\setcoordinatesystem units <0.50truein,0.50truein>
\setplotarea x from -2.5 to 3.0, y from -0.25 to 3.0
\plot -2.00 1.54  -2.00 1.16  0.00 0.00  2.75 1.16  1.91 1.30 /
\plot 0.00 0.00  0.00 0.75  2.75 1.56  0.56 1.85  -2.00 1.54 
0.00 0.75  /
\plot 2.75 1.16  2.75 1.56 /
\setquadratic
\plot -2.00 2.824  -1.75 2.570  -1.50 2.379  -1.25 2.237
-1.00 2.133  -0.75 2.059  -0.50 2.008  -0.25 1.975
0.00 1.9560  0.25 1.9478  0.56 1.9493  /
\setlinear
\arrow <5pt> [0.2,0.67] from 0.00 -0.25 to 0.75 0.09
\arrow <5pt> [0.2,0.67] from 0.00 -0.25 to -0.50 0.10
\arrow <5pt> [0.2,0.67] from -2.20 1.16 to -2.20 1.54
\setshadegrid span <1.5pt>
\vshade -2.00 1.16 1.54  <,,z,z>  0.00 0.00 0.75 /
\vshade 1.91 1.30 1.32  <,,z,z>  2.75 1.16 1.56 /
\setdashes
\setdashpattern <2pt, 2pt>
\plot -2.00 1.54  -2.00 3.00 /
\plot 0.56 1.86  0.56 2.00 /
\put {{\ninepoint $y$}} [r] at -2.30 1.35
\put {{\ninepoint $r$}} [l] at 0.50 -0.25
\put {{\ninepoint $x^\mu$}} [r] at -0.40 -0.20
\put {{\ninepoint $e^{-2|r|/\ell}$}} [c] at -0.75 2.50
\put {{\ninepoint Planck brane}} [r] at -1.25 0.50
\put {{\ninepoint TeV brane}} [l] at 3.00 1.36
\endpicture$$
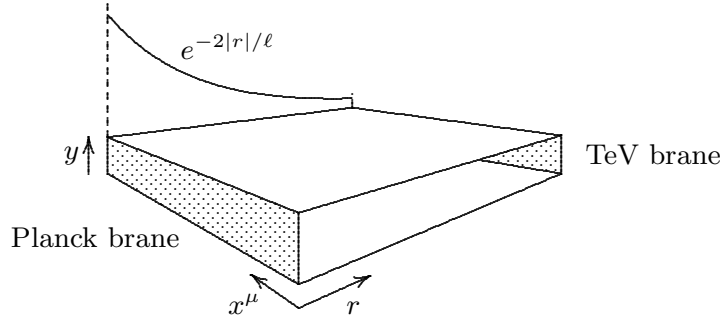
{\ninepoint Figure 1.  The geometry of a six-dimensional model with two
$4$-branes.  The small periodic coordinate is $y$.  The direction orthogonal
to the $4$-branes, $r$, becomes the extra coordinate of the Randall-Sundrum
model when we integrate out the $y$ dimension.  We can recover the second
Randall-Sundrum model $\rsb$ by letting $r_c\to\infty$.  The model assumes an
orbifold geometry about $r=0$.}
\smallskip

The existence of metric solutions that are periodic in $y$ appears to be a
fairly generic feature of actions that include terms beyond the standard
Einstein-Hilbert terms $\warped$.  For example, when the theory is truncated
so that the gravitational action contains only terms with up to four
derivatives of the metric and including a scalar field $\phi$ with
$\Delta{\cal L} = k (\nabla_a\phi \nabla^a\phi)^2$, the equations of motion
for $\action$ admit periodic solutions over a range of the parameters, $\{
\Lambda, a, b, c, k\}$.  

We instead shall focus upon the simpler case where $\Delta{\cal L}$ describes
a Casimir effect.  Since the finite $y$-direction explicitly breaks $5+1$
dimensional Poincar\' e symmetry, any quantum vacuum contribution can differ
in the $y$ and $(x^\lambda,r)$ directions.  We can include this effect by
adding the following energy-momentum tensor to the field equations,
$$\hbox{$T^{\rm vac}$}^{\ b}_{a} = {\rm diag}\left( C, C, C, C, C, -C \right)
.  \neqno\casimirdef$$
Any contribution proportional to the identity can be absorbed into the
definition of the cosmological constant.\foot{Since we do not require that
$\Lambda$ vanish, quantum contributions no longer need to cancel the classical
contribution to $\Lambda$; rather, they are only regarded as some component of
the total $\Lambda$.}  Satisfactory, periodic solutions then exist when $C$ is
above a mild bound.

To produce a Randall-Sundrum scenario in five of the dimensions, we shall
examine a metric of the form
$$ds^2 = g_{ab}(x^\lambda,r,y)\, dx^adx^b 
= e^{A(y)} \hat g_{MN}(x^\lambda,r)\, dx^M dx^N + dy^2 \neqno\bulkmetric$$
with an AdS$_5$ metric for the $(x^\lambda, r)$-subspace,
$$d\hat s^2 = \hat g_{MN}\, dx^M dx^N = e^{-2|r|/\ell}\eta_{\mu\nu}\, dx^\mu
dx^\nu + dr^2 . \neqno\AdSmetric$$
When $A(y)$ is a periodic function of $y$, we can obtain a compact extra
dimension with a very non-trivial $y$-dependence without any singularities. 
The shape of $A(y)$ determines the effective cosmological constant of the
$\hat g_{MN}$ metric.  Since this metric is conformally flat, the linear
combinations of $R^2$ terms that represents the squared Weyl tensor will not
contribute to the equations of motion.  It is convenient to parameterize the
two remaining linear combinations by
$$\tilde\mu \equiv 20a+6b+4c \qquad \tilde\lambda \equiv
15a+{\textstyle{5\over 2}}b+c . \neqno\paradefs$$

Before examining the detailed form of the equations that determine the warp
function $A(y)$ with a Randall-Sundrum $5d$ subspace, we derive a $5d$
effective action by integrating out the sixth dimension in the warped
background given by the metric $\bulkmetric$.  There, the scalar curvature $R$
is related to the scalar curvature $\hat R$ of the $5d$ metric $\hat g_{MN}$
and derivatives of the warp function,
$$R = e^{-A(y)} \hat R - 5\, A^{\prime\prime} - {\textstyle{15\over 2}}\,
(A')^2 . \neqno\RScomps$$
Here the prime denotes a $y$-derivative.  Similarly, the components of the
Ricci and Riemann tensors in the full theory, $R_{ab}$ and $R^a_{\ bcd}$, can
be expanded in terms of the analogous tensors for the Randall-Sundrum
subspace, $\hat R_{MN}$ and $\hat R_{MNPQ}$.

Integrating out the small $y_c$ dimension in a background such as $\AdSmetric$
where $\hat R$ is constant produces a five dimensional effective action,
$$\eqalign{ S^{\rm eff}_{\rm bulk} 
&= M_5^3 \int d^4xdr\, \sqrt{-\hat g}\, \Bigl( 2\Lambda_{\rm eff} + \hat R +
a_{\rm eff} \hat R^2 + b_{\rm eff} \hat R_{MN}\hat R^{MN} \cr 
&\quad \qquad\qquad\qquad\qquad 
+ c_{\rm eff} \hat R_{MNPQ}\hat R^{MNPQ} + \cdots \Bigr) . \cr}
\neqno\effectaction$$
This action should reproduce the leading behavior for small $x^\mu$-dependent
perturbations about AdS$_5$ $\AdSmetric$, ${\bf R}^{4,1}$ or de Sitter space
backgrounds.  The new parameters that appear in this effective action depend
partially upon the ``fundamental'' parameters of the original action but also
upon the behavior of the warp function.  Thus, in the low energy theory the
$5d$ cosmological constant is
$$M_5^3 \Lambda_{\rm eff} = M_6^4\int_0^{y_c} dy\, e^{{5\over 2}A(y)} \left[
\Lambda - {\textstyle{1\over 4}} (\phi')^2 + {\textstyle{5\over 2}} (A')^2 +
{\textstyle{5\over 8}}\tilde\mu (A^{\prime\prime})^2 - {\textstyle{5\over 24}}
\tilde\lambda (A')^4 \right] \neqno\effparaa$$
while the $5d$ Planck mass is
$$M_5^3 = M_6^4 \int_0^{y_c} dy\, e^{{3\over 2}A(y)} \left[ 1 
- {\textstyle{1\over 8}} (3\tilde\mu - 4\tilde\lambda) (A')^2 \right] .
\neqno\effparab$$
The coefficients of the $\hat R^2$ terms are 
$$M_5^3 a_{\rm eff} = M_6^4 a \int_0^{y_c} dy\, e^{{1\over 2}A(y)} ,
\neqno\effparac$$
with analogous expressions for $b_{\rm eff}$ and $c_{\rm eff}$.  In these
expressions we have freely integrated by parts.

For the theory to resemble the standard Randall-Sundrum picture, the five
dimensional theory of gravity should be weak, $M_5\ell\gg 1$.  Since
$\Lambda_{\rm eff}\sim \ell^{-2}$, we require the effective $5d$ cosmological
constant to be small which can easily occur when the contribution from the
bulk cosmological constant is partially cancelled by effects from the warp
function in $\effparaa$.  In the appendix, we show two such examples of
periodic solutions with $\Lambda_{\rm eff}\not=0$ but small.

In the weak $5d$ gravity limit, $M_5\ell\gg 1$, the self-coupling terms become
negligible and the leading behavior is governed by the Einstein-Hilbert terms
in $\effectaction$.  If we include two $4$-branes at $r=0$ and $r=r_c$ with
respective tensions $\sigma^{(0)}$ and $\sigma^{(r_c)}$, then we recover the
action considered by Randall and Sundrum $\rsa$,
$$\eqalign{ S^{\rm eff}
&= M_5^3 \int d^4xdr\, \sqrt{-\hat g}\, \Bigl( 2\Lambda_{\rm eff} + \hat R
\Bigr) \cr 
&\quad 
+ M_5^3 \int_{r=0} d^4x\, \sqrt{-\hat h}\, \left[ 2\sigma_{\rm eff}^{(0)}
\right] + M_5^3 \int_{r=r_c} d^4x\, \sqrt{-\hat h}\, \left[ 2\sigma_{\rm
eff}^{(r_c)} \right]
 + \cdots .  \cr} \neqno\effaction$$
Here the effective tension on the $r=0$ brane is
$$M_5^3 \sigma^{(0)}_{\rm eff}
= M_6^4 \sigma^{(0)} \int_0^{y_c} dy\, e^{2A(y)} , \neqno\effparad$$
with an analogous expression for the $r=r_c$ brane.  $\hat h_{MN}$ represents
the metric induced on the branes at $r=0$ or $r=r_c$ by the metric $\hat
g_{MN}$.  

The fine-tunings of the tensions on the Planck brane ($\sigma_{\rm
eff}^{(0)}$) and the TeV brane ($\sigma_{\rm eff}^{(r_c)}$) in $\rsa$ are 
$$\Lambda_{\rm eff} = - 6 \sigma_{\rm eff}^{(0)} = 6 \sigma_{\rm eff}^{(r_c)}.
\neqno\finetune$$
Goldberger and Wise 
\ref\gw{W.~D.~Goldberger and M.~B.~Wise, ``Modulus stabilization with bulk
fields,'' Phys.\ Rev.\ Lett.\ {\bf 83}, 4922 (1999) [hep-ph/9907447].} 
showed that including a massive bulk scalar field with quartic couplings to
the brane, thereby generating a non-trivial effective potential for $r_c$,
removes one of the fine-tunings inherent in $\finetune$.  Since we expect that
some such mechanism can be adapted to our picture, we are left with one
condition on $\Lambda_{\rm eff}$.  However, $\Lambda_{\rm eff}$ is itself a
derived parameter and solutions of the equations of motion from $\action$ can
satisfy $\finetune$ without any unnatural choices for the parameters in the
action.  The vanishing of the $4d$ effective cosmological constant in the
Randall-Sundrum scenario in this picture thus reduces to a dynamical question
as to why the flat solutions are favored.

\newsec{Exact analysis.}

We have asserted that the equations of motion for the $6d$ action admit
solutions for which $A(y)$ is periodic with an AdS$_5$ (or a dS$_5$) metric in
the remaining dimensions.  The method for numerically demonstrating this claim
has been detailed in $\warped$ for a $5d$ dimensional theory with a flat $4d$
subspace.  The qualitative results do not change when we start from a theory
in $5+1$ dimensions and alter only slightly when a small cosmological constant
appears in the $5d$ subspace.  The flat solutions, for which $\hat
g_{MN}=\eta_{MN}$ in $\bulkmetric$, exist in the region of parameter space
where
$$\Lambda,\tilde\mu < 0 \quad\hbox{and}\quad 
\tilde\lambda\Lambda > \cases{ 
-{25\over 4}\tilde\mu\Lambda + 5 \sqrt{10\tilde\mu\Lambda} - {15\over 2}
       &for $0< \tilde\mu\Lambda \le {9\over 10}$\cr
{25\over 12}\tilde\mu\Lambda  
       &for ${9\over 10} \le \tilde\mu\Lambda$\cr}
\neqno\paraspace$$
although the exact location of the boundary corresponding to the second case
of $\paraspace$ has not been precisely determined.

When a $5d$ cosmological constant $\Lambda_{\rm eff}$ is included, the form of
$A(y)$ alters slightly from its $\Lambda_{\rm eff}=0$ value; however, the warp
function remains periodic as long as $|\Lambda_{\rm eff}| \mathop{\roughly{<}}
{\cal O}(|\Lambda|)$.  This bound includes the case wherein the $5d$ effective
theory of gravity is weak.  Thus it is possible to satisfy $M_5^2\Lambda_{\rm
eff}^{-1}\gg 1$ even while $M_6^2\Lambda^{-1}\sim 1$.

The bulk equations of motion for a universe with an AdS$_5$ subspace
$\AdSmetric$, are obtained by varying the full action $\action$,
$$\eqalign{
\tilde\mu \left[ {\textstyle{1\over 2}} A^{\prime\prime\prime\prime} 
+ {\textstyle{5\over 2}} A'A^{\prime\prime\prime}
+ {\textstyle{15\over 8}} (A^{\prime\prime})^2
+ {\textstyle{25\over 8}} A^{\prime\prime}(A')^2 \right] & \cr
+ \tilde\lambda \left[ A^{\prime\prime}(A')^2
+ {\textstyle{5\over 8}} (A')^4 \right]
- 2 A^{\prime\prime} - {\textstyle{5\over 2}} (A')^2 & \cr
- {\textstyle{1\over\ell^2}} e^{-A(y)} \left[ 6 
+ ( 3 \tilde\mu - 4 \tilde\lambda ) 
\left[ A^{\prime\prime} + {\textstyle{3\over 4}} (A')^2 \right] \right] & \cr
+ {\textstyle{2\over\ell^4}} e^{-2A(y)} \left[ {\textstyle{1\over 4}}
\tilde\mu + \tilde\lambda \right]  
&= - \Lambda - C + {\textstyle{1\over 4}} (\phi')^2 \cr
{\textstyle{5\over 4}} \tilde\mu \left[ A'A^{\prime\prime\prime}
- {\textstyle{1\over 2}} (A^{\prime\prime})^2
+ {\textstyle{5\over 2}} A^{\prime\prime}(A')^2 \right] 
+ {\textstyle{5\over 8}} \tilde\lambda (A')^4 
- {\textstyle{5\over 2}} (A')^2 & \cr
- {\textstyle{1\over\ell^2}} e^{-A(y)} \left[ 10 
+ {\textstyle{5\over 4}} ( 3 \tilde\mu - 4 \tilde\lambda ) (A')^2 \right] 
+ {\textstyle{10\over\ell^4}} e^{-2A(y)} \left[ {\textstyle{1\over 4}}
\tilde\mu + \tilde\lambda \right]  
&= - \Lambda + C - {\textstyle{1\over 4}} (\phi')^2 . \cr} 
\neqno\eqnofmotion$$
The equation of motion for the scalar field is
$$\phi^{\prime\prime} + {\textstyle{5\over 2}} A' \phi' = 0 \neqno\phieqn$$
when the scalar field depends only on the $y$-direction, $\phi=\phi(y)$.  The
sum of the equations in $\eqnofmotion$ provide a single differential equation
for $A(y)$ which we can integrate numerically.  Once a periodic solution is
found, the difference of these equations determines the behavior of the scalar
field.  Note that since the scalar field appears in $\eqnofmotion$ and
$\phieqn$ only through its derivatives, only $\phi'$ is guaranteed to be
periodic; $\phi(y)$ monotonically increases and must thus assume values only
over a compact range.

We have found periodic solutions throughout the parameter space $\paraspace$
for small $5d$ cosmological constants of either sign.  Two such examples are
sketched in the appendix.  Since the theory is invariant under
$y$-translations and rescalings in the other five coordinates, we are free to
choose coordinates in which $A(0)=A'(0)=0$ without any loss of generality.  We
also set $A^{\prime\prime\prime}(0)=0$.  We then numerically integrated
$\eqnofmotion$ for an initial choice for $A^{\prime\prime}(0)$.  What we find
is that for a particular value of $A^{\prime\prime}(0)$, the warp function
returns to its initial conditions after some finite $y_c$:  $A(y)=A(y+y_c)$.

Note that when we substitute the equations of motion $\eqnofmotion$ into the
expression for the effective cosmological constant $\effparaa$,
$$M_5^3 \Lambda_{\rm eff} = M_6^4\int_0^{y_c} dy\, e^{{3\over 2}A(y)} \left[
{\textstyle{6\over\ell^2}} - {\textstyle{6\over\ell^2}} (b+c) (A')^2 -
{\textstyle{4\over\ell^4}} e^{-A(y)} (10a+2b+c)  \right] , \neqno\effAdS$$
and rewrite this expression in terms of the parameters of the effective theory 
$\effparaa$--$\effparac$, we find
$$\Lambda_{\rm eff} = {6\over\ell^2}  - {4\over\ell^4} (10a_{\rm eff}+2b_{\rm
eff}+c_{\rm eff}) = {6\over\ell^2}  - {8\over\ell^4} \lambda_{\rm eff} . 
\neqno\effAdSLambda$$
We would have obtained the same result by inserting the metric in equation
$\AdSmetric$ into the equations of motion for the effective action,
$\effectaction$.

\newsec{Thick branes.}

In passing to the weak gravity limit in order to ignore the $\hat R^2$ terms
in the $5d$ theory $\effaction$, we might worry that while their effect on the
bulk dynamics is small, they nevertheless might significantly alter the brane
tension.  Indeed, the particular linear combination of terms for which 
$$\mu_{\rm eff} \equiv 16 a_{\rm eff} + 5 b_{\rm eff} + 4 c_{\rm eff} \not= 0
\neqno\effmu$$
generates terms of the form $\partial_r^2\delta(r)$ or $(\delta(r))^2$ for
$\AdSmetric$ which implies that the brane tension receives an infinite
correction.  Note that this feature is present in the original Randall-Sundrum
scenario.  Therefore, in its original form, even in the weak field limit and
when the coefficients of the $\hat R^2$ terms are small but generic, small
effects in the bulk can have a large effect on the brane tension.

The origin for these $\delta$-function divergences is the zero thickness of
the brane.  Yet while we can remove such severe divergences by giving the
brane a finite thickness, in order to recover the Randall-Sundrum picture we
must additionally show that the $\hat R^2$ terms can be neglected in this
case.  Therefore, we shall examine a solution of the form,
$$\hat g_{MN}\, dx^m dx^N = e^{\sigma(r)} \eta_{\mu\nu}\, dx^\mu dx^\nu +
dr^2, \neqno\thickmetric$$
for the exact $a_{\rm eff} = b_{\rm eff} = c_{\rm eff} = 0$ case, and then
study the size of $\hat R$ compared to $M_5^2$ both near the brane and deeply
within the bulk.

An elegant formalism for obtaining a thick domain wall generated by a scalar
field $\Phi(r)$,
$$S = M_5^3 \int d^4xdr\, \sqrt{-\hat g} \left[ \hat R - {\textstyle{1\over
2}} \hat\nabla_M\Phi \hat\nabla^M\Phi - V(\Phi) \right] , \neqno\wallaction$$
has been used by several authors 
\ref\cvetic{M.~Cveti\v c, S.~Griffies and S.~Rey, ``Static domain walls in
$N=1$ supergravity,'' Nucl.\ Phys.\ B {\bf 381}, 301 (1992) [hep-th/9201007].} 
\ref\thickbrane{K.~Skenderis and P.~K.~Townsend, ``Gravitational stability and
renormalization-group flow,'' Phys.\ Lett.\ B {\bf 468}, 46 (1999)
[hep-th/9909070] and
O.~DeWolfe, D.~Z.~Freedman, S.~S.~Gubser and A.~Karch, ``Modeling the fifth
dimension with scalars and gravity,'' Phys.\ Rev.\ D {\bf 62}, 046008 (2000)
[hep-th/9909134].} 
\ref\gremm{M.~Gremm, ``Four-dimensional gravity on a thick domain wall,''
Phys.\ Lett.\ B {\bf 478}, 434 (2000) [hep-th/9912060].} 
who noted that when the scalar potential has the following `superpotential'
form,
$$V(\Phi) = {1\over 2} \left( {\partial W\over\partial\Phi} \right)^2 -
{1\over 3} W^2(\Phi) , \neqno\superpot$$
the $5d$ warp function and domain wall profile are respectively,
$$\partial_r\sigma = -{1\over 3} W(\Phi) \qquad 
\partial_r\Phi = {\partial W\over\partial\Phi} . \neqno\bpssoln$$

We shall use the solution of Gremm $\gremm$ for a single domain wall at $r=0$
($r_c\to\infty$) with $W(\Phi) = {6\over\ell} \sin\left( \sqrt{
{6\over\kappa\ell}} \Phi \right)$ and
$$\eqalign{
\sigma(r) &= -{\textstyle{2\over\kappa\ell}} \ln\left[ 2\cosh(\kappa r)
\right] \cr
\Phi(r)   &= 2\sqrt{{\textstyle{6\over\kappa\ell}}} \arctan\left[
\tanh({\textstyle{1\over 2}} \kappa r) \right] . \cr} \neqno\gremmsoln$$
Here $\ell$ corresponds to the asymptotic AdS$_5$ length as in $\AdSmetric$
and $\kappa^{-1}$ is the thickness of the brane.  The scalar curvature is now
free of singularities everywhere,
$$\hat R = {8\kappa\over\ell} - \left( {8\kappa\over\ell} + {20\over\ell^2}
\right) \tanh^2(\kappa r) . \neqno\hatReg$$
From this result we notice that in the bulk ($r\gg\kappa^{-1}$), since $\hat R
\to -20\ell^{-2}$, the weak gravity condition is $M_5\ell\gg 1$.
At the brane ($r\approx 0$), in order to be able to neglect the $\hat R^2$
terms relative to $\hat R$, we require
$$M_5 \sqrt{\ell\over\kappa} \gg 1 . \neqno\weakatbrane$$
Thus, assuming that gravity is weakly coupled in the bulk so that $M_5\ell\gg
1$, both $\kappa\sim\ell^{-1}$ and $\kappa\sim M_5$ automatically satisfy
$\weakatbrane$.  

The above analysis applies to the positive tension brane, or `Planck brane'
$\rsa$, although the same infinite corrections to the negative tension brane
arise when $\mu_{\rm eff}\not =0$.  Negative tension branes do not admit a
thick wall description, at least in the regime in which the $\hat R^2$ terms
become negligible.  However, we can also add a warped, compact extra dimension
to a scenario that does not contain any thin negative tension branes, such as
that of Lykken and Randall 
\ref\lr{J.~Lykken and L.~Randall, ``The shape of gravity,'' JHEP{\bf 0006},
014 (2000) [hep-th/9908076].}.  

The purely gravitational action of $\effectaction$, which includes $\hat R^2$
terms, can also generate a warp function in the $5d$ subspace $\thickmetric$
that resembles that in $\gremmsoln$ but without the need for a scalar field to
generate a domain wall.  In this case, we still have\foot{This configuration
corresponds to equation $(4.6)$ of $\warped$.} 
$$\sigma(r) = - {\textstyle{2\over\kappa\ell}} \ln\left[ 2\cosh(\kappa r)
\right] , \neqno\pseudowarp$$
but where the width, $\kappa^{-1}$, and the asymptotic AdS$_5$ length, $\ell$,
are respectively
$$\kappa = \left( {3-4\sqrt{2\Lambda_{\rm eff}\mu_{\rm eff}}\over -2\mu_{\rm
eff}} \right)^{1/2}
\qquad
\ell = \left( {3-4\sqrt{2\Lambda_{\rm eff}\mu_{\rm eff}}\over -\Lambda_{\rm
eff}} \right)^{1/2} , \neqno\pseudowall$$
with $0 \le \Lambda_{\rm eff}\mu_{\rm eff} \le {9\over 32}$, $\Lambda_{\rm
eff}<0$ and $\mu_{\rm eff} \le 0$.  This configuration requires one
fine-tuning among $\Lambda_{\rm eff}$, $\lambda_{\rm eff}$ and $\mu_{\rm
eff}$, given in $\warped$ by 
$$\Lambda_{\rm eff}\lambda_{\rm eff} 
= - \left( 3 - 4\sqrt{2\Lambda_{\rm eff}\mu_{\rm eff}} \right) 
\left( {\textstyle{9\over 8}} - {\textstyle{1\over 2}} \sqrt{2\Lambda_{\rm
eff}\mu_{\rm eff}} \right) , \neqno\littlelambda$$
which can presumably be effected by the appropriately warped compactification
in the sixth dimension.  

The $\hat R^2$ terms in this case are in no sense negligible---they play the
same role as $\Phi(r)$ above and balance against the $\hat R$ term to produce
the solution $\pseudowarp$.  In particular, $|\hat R| \gg |\lambda_{\rm
eff}\hat R^2|$ for $r\gg \kappa^{-1}$ translates into
$${\textstyle{45\over 2}} - 10 \sqrt{2\Lambda_{\rm eff}\mu_{\rm eff}} \ll 1
\neqno\lambdaconstraint$$
which is nowhere satisfied in the allowed range for $\Lambda_{\rm eff}\mu_{\rm
eff}$.

\newsec{Concluding remarks.}

An intriguing feature of this model is that for each choice of the parameters
in the action that admits a periodic warp function, a family of solutions
exists.  Elements of this family are specified by the value of the $5d$
effective cosmological constant, $\Lambda_{\rm eff}$.  Alternatively, since
each $\Lambda_{\rm eff}$ is associated with a unique period---or at worst a
discrete set of periods---of the warp function, we can also specify an element
by the size of the sixth dimension, $y_c$.  This behavior differs from a
factorizable geometry such as ${\bf R}^{3,1}\times S^1$ or AdS$_4\times S^1$
which has a continuous set of solutions for each $\Lambda_{\rm eff}$---now a
fundamental parameter in the action---labeled by a compactification radius
which is not fixed by the action.

While this scenario does not require any unnatural choices of the parameters
in $\action$ to satisfy $\finetune$, this condition is not the unique solution
to the equations of motion.  Some further dynamical mechanism is still
required that favors a bulk $\Lambda_{\rm eff}$ that obeys $\finetune$.  

The picture that we have described allows solutions with a stable exponential
electroweak-Planck hierarchy without an unnatural choice of the parameters in
the action.  Since this picture crucially relies on the presence of $R^2$
terms in the action, it might be worried whether it persists upon including
further higher order $R^n$ terms.  However, in $\warped$ we argued that such
periodic solutions should exist generically for an action composed of general
powers of the curvature tensors.  In particular, the small cosmological
constant case, $M_6^2\Lambda^{-1}\ll 1$, with $\Lambda_{\rm eff}=0$ admitted a
semi-analytic description of these periodic solutions.  Including a small $5d$
cosmological constant does not greatly perturb these solutions, so that the
physically interesting case in which the effective $5d$ theory of gravity is
weak should continue to exist even with higher order terms in the underlying
six dimensional theory.

\bigskip\bigskip
\centerline{ {\bf Acknowledgements} }
\medskip

\noindent
This work was supported in part by the Natural Sciences and Engineering
Research Council of Canada.

\appendix{A}{Examples of periodic warp functions.}

Since the existence of periodic metrics resulting from $\eqnofmotion$ can only
be demonstrated numerically, it is helpful to present a couple of examples. 
For a small $\Lambda_{\rm eff}$, that is $M_6\ell\gg 1$, the shape of the warp
function $A(y)$ and the parameter space in which periodic ones exist
$\paraspace$ are nearly the same as in the $\Lambda_{\rm eff}=0$ case.  In
figures 2 and 3, the AdS$_5$ length $\ell$ that appears in the $\hat g_{MN}$
components of the metric is the same in both cases although the forms of
$A(y)$ differ markedly in the two examples.  Figure 2 shows a typical profile
of a warp function for a small cosmological constant, $M_6^2\Lambda^{-1}\gg
1$, while figure 3 shows an example in which $M_6^2\Lambda^{-1}\sim 1$.  
$$\beginpicture
\setcoordinatesystem units <1.0truein,1.0truein>
\setplotarea x from 0.0 to 3.0, y from -0.1 to 1.0
\setlinear
\plot  0.000 0.00  .015 .0023  .030 .0092  .045 .0207  .060 .0365  .075 .0566
.090 .0805  .105 .1082  .120 .1392  .135 .1733  .150 .2099  .165 .2488
.180 .2895  .195 .3315  .210 .3745  .225 .4179  .240 .4614  .255 .5045
.270 .5468  .285 .5879  .300 .6275  .315 .6651  .330 .7004  .345 .7332
.360 .7631  .375 .7899  .390 .8133  .405 .8332  .420 .8495  .435 .8619
.450 .8704  .465 .8749  .480 .8754  .495 .8718  .510 .8643  .525 .8528
.540 .8375  .555 .8184  .570 .7958  .585 .7698  .600 .7406  .615 .7085
.630 .6738  .645 .6367  .660 .5976  .675 .5568  .690 .5148  .705 .4718
.720 .4284  .735 .3849  .750 .3418  .765 .2995  .780 .2585  .795 .2191
.810 .1819  .825 .1472  .840 .1154  .855 .0869  .870 .0620  .885 .0410
.900 .0241  .915 .0116  .930 .0036  .945 .0001  .960 .0013  .975 .0072
.990 .0175  1.005 .0323  1.020 .0514  1.035 .0744  1.050 .1012  1.065 .1315
1.080 .1648  1.095 .2009  1.110 .2393  1.125 .2796  1.140 .3213  1.155 .3641
1.170 .4074  1.185 .4510  1.200 .4942  1.215 .5367  1.230 .5782  1.245 .6181
1.260 .6562  1.275 .6921  1.290 .7255  1.305 .7561  1.320 .7837  1.335 .8080
1.350 .8288  1.365 .8459  1.380 .8593  1.395 .8687  1.410 .8742  1.425 .8756
1.440 .8731  1.455 .8665  1.470 .8559  1.485 .8415  1.500 .8233  1.515 .8016
1.530 .7764  1.545 .7479  1.560 .7165  1.575 .6824  1.590 .6458  1.605 .6072
1.620 .5668  1.635 .5250  1.650 .4822  1.665 .4389  1.680 .3954  1.695 .3521
1.710 .3096  1.725 .2682  1.740 .2284  1.755 .1906  1.770 .1553  1.785 .1227
1.800 .0934  1.815 .0676  1.830 .0457  1.845 .0278  1.860 .0142  1.875 .0051
1.890 .0005  1.905 .0006  1.920 .0053  1.935 .0146  1.950 .0284  1.965 .0464
1.980 .0685  1.995 .0945  2.010 .1239  2.025 .1565  2.040 .1920  2.055 .2299
2.070 .2697  2.085 .3112  2.100 .3537  2.115 .3970  2.130 .4405  2.145 .4838
2.160 .5266  2.175 .5683  2.190 .6087  2.205 .6472  2.220 .6837  2.235 .7177
2.250 .7491  2.265 .7774  2.280 .8025  2.295 .8241  2.310 .8421  2.325 .8564
2.340 .8668  2.355 .8732  2.370 .8756  2.385 .8740  2.400 .8684  2.415 .8588
2.430 .8453  2.445 .8281  2.460 .8071  2.475 .7827  2.490 .7551  2.505 .7243
2.520 .6908  2.535 .6548  2.550 .6167  2.565 .5766  2.580 .5352  2.595 .4926
2.610 .4493  2.625 .4058  2.640 .3625  2.655 .3197  2.670 .2780  2.685 .2378
2.700 .1995  2.715 .1635  2.730 .1303  2.745 .1002  2.760 .0735  2.775 .0506
2.790 .0317  2.805 .0171  2.820 .0069  2.835 .0012  2.850 .0002  2.865 .0038
2.880 .0120  2.895 .0247  2.910 .0417  2.925 .0628  2.940 .0879  2.955 .1165
2.970 .1484  2.985 .1832  3.000 .2205  /
\putrule from -0.1 0.0 to 3.1 0.0
\putrule from 0.0 -0.1 to 0.0 1.0
\putrule from 0.3 -0.02 to 0.3 0.02
\putrule from 0.6 -0.02 to 0.6 0.02
\putrule from 0.9 -0.02 to 0.9 0.02
\putrule from 1.2 -0.02 to 1.2 0.02
\putrule from 1.5 -0.035 to 1.5 0.035
\putrule from 1.8 -0.02 to 1.8 0.02
\putrule from 2.1 -0.02 to 2.1 0.02
\putrule from 2.4 -0.02 to 2.4 0.02
\putrule from 2.7 -0.02 to 2.7 0.02
\putrule from 3.0 -0.035 to 3.0 0.035
\putrule from -0.035 1.0 to 0.035 1.0
\putrule from -0.035 0.5 to 0.035 0.5
\put {$A(y)$} [c] at 0.20 1.0 
\put {$y$} [c] at 3.15 0.1 
\put {${\scriptstyle 0}$} [c] at -0.15 0 
\put {${\scriptstyle 1}$} [c] at 1.5 -0.15 
\put {${\scriptstyle 2}$} [c] at 3.0 -0.15
\put {${\scriptstyle 0.1}$} [c] at -0.15 1.00 
\endpicture$$
{\ninepoint  \baselineskip=10pt
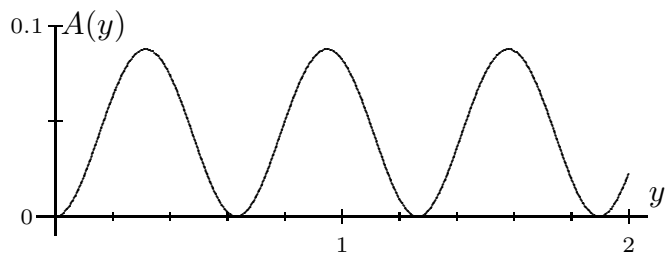
{\bf Figure 2.\/}  A periodic warp function $A(y)$ for $\Lambda=-1$ and
$\lambda=\mu=-0.1$.  The initial condition is $A^{\prime\prime}(0) =
4.63923991$.  The value of the AdS length which appears in $\AdSmetric$ is
$\ell=10$.}
\medskip

$$\beginpicture
\setcoordinatesystem units <1.0truein,1.0truein>
\setplotarea x from 0.0 to 3.0, y from -0.1 to 1.0
\setlinear
\plot  0.00 0.00  .015 .006  .030 .023  .045 .049  .060 .084  .075 .125
.090 .170  .105 .217  .120 .266  .135 .313  .150 .360  .165 .404
.180 .446  .195 .484  .210 .520  .225 .553  .240 .582  .255 .609
.270 .633  .285 .654  .300 .673  .315 .689  .330 .703  .345 .715
.360 .726  .375 .735  .390 .742  .405 .748  .420 .752  .435 .756
.450 .758  .465 .760  .480 .760  .495 .760  .510 .759  .525 .756
.540 .753  .555 .749  .570 .743  .585 .736  .600 .728  .615 .718
.630 .706  .645 .692  .660 .676  .675 .658  .690 .637  .705 .614
.720 .588  .735 .559  .750 .527  .765 .492  .780 .454  .795 .413
.810 .369  .825 .323  .840 .275  .855 .227  .870 .179  .885 .134
.900 .092  .915 .056  .930 .027  .945 .008  .960 .000  .975 .004
.990 .018  1.005 .04  1.020 .076  1.035 .116  1.050 .161  1.065 .208
1.080 .256  1.095 .304  1.110 .351  1.125 .395  1.140 .437  1.155 .477
1.170 .513  1.185 .546  1.200 .577  1.215 .604  1.230 .628  1.245 .650
1.260 .669  1.275 .686  1.290 .701  1.305 .713  1.320 .724  1.335 .733
1.350 .740  1.365 .747  1.380 .751  1.395 .755  1.410 .758  1.425 .759
1.440 .760  1.455 .760  1.470 .759  1.485 .757  1.500 .754  1.515 .750
1.530 .744  1.545 .738  1.560 .730  1.575 .720  1.590 .708  1.605 .695
1.620 .680  1.635 .662  1.650 .642  1.665 .619  1.680 .593  1.695 .565
1.710 .534  1.725 .499  1.740 .461  1.755 .421  1.770 .378  1.785 .332
1.800 .285  1.815 .237  1.830 .189  1.845 .143  1.860 .100  1.875 .062
1.890 .032  1.905 .011  1.920 .001  1.935 .002  1.950 .015  1.965 .037
1.980 .069  1.995 .108  2.010 .152  2.025 .198  2.040 .246  2.055 .294
2.070 .341  2.085 .387  2.100 .429  2.115 .469  2.130 .506  2.145 .540
2.160 .571  2.175 .599  2.190 .624  2.205 .646  2.220 .666  2.235 .683
2.250 .698  2.265 .711  2.280 .722  2.295 .731  2.310 .739  2.325 .745
2.340 .750  2.355 .754  2.370 .757  2.385 .759  2.400 .760  2.415 .760
2.430 .759  2.445 .757  2.460 .754  2.475 .750  2.490 .745  2.505 .739
2.520 .731  2.535 .722  2.550 .711  2.565 .698  2.580 .683  2.595 .666
2.610 .646  2.625 .624  2.640 .599  2.655 .571  2.670 .540  2.685 .506
2.700 .469  2.715 .429  2.730 .387  2.745 .342  2.760 .295  2.775 .246
2.790 .198  2.805 .152  2.820 .108  2.835 .069  2.850 .037  2.865 .015
2.880 .002  2.895 .001  2.910 .011  2.925 .032  2.940 .062  2.955 .096
2.970 .142  2.985 .189  3.000 .237  /
\putrule from -0.1 0.0 to 3.1 0.0
\putrule from 0.0 -0.1 to 0.0 1.0
\putrule from 0.25 -0.02 to 0.25 0.02
\putrule from 0.50 -0.02 to 0.50 0.02
\putrule from 0.75 -0.02 to 0.75 0.02
\putrule from 1.00 -0.02 to 1.00 0.02
\putrule from 1.25 -0.02 to 1.25 0.02
\putrule from 1.50 -0.02 to 1.50 0.02
\putrule from 1.75 -0.02 to 1.75 0.02
\putrule from 2.00 -0.02 to 2.00 0.02
\putrule from 2.25 -0.02 to 2.25 0.02
\putrule from 2.50 -0.02 to 2.50 0.02
\putrule from 2.75 -0.02 to 2.75 0.02
\putrule from 3.00 -0.02 to 3.00 0.02
\putrule from -0.035 1.0 to 0.035 1.0
\putrule from -0.035 0.5 to 0.035 0.5
\put {$A(y)$} [c] at 0.20 1.0 
\put {$y$} [c] at 3.15 0.1 
\put {${\scriptstyle 0}$} [c] at -0.15 0 
\put {${\scriptstyle 1}$} [c] at 0.25 -0.1
\put {${\scriptstyle 2}$} [c] at 0.50 -0.1
\put {${\scriptstyle 3}$} [c] at 0.75 -0.1
\put {${\scriptstyle 4}$} [c] at 1.00 -0.1
\put {${\scriptstyle 5}$} [c] at 1.25 -0.1
\put {${\scriptstyle 6}$} [c] at 1.50 -0.1
\put {${\scriptstyle 7}$} [c] at 1.75 -0.1
\put {${\scriptstyle 8}$} [c] at 2.00 -0.1
\put {${\scriptstyle 9}$} [c] at 2.25 -0.1
\put {${\scriptstyle 10}$} [c] at 2.50 -0.1
\put {${\scriptstyle 11}$} [c] at 2.75 -0.1
\put {${\scriptstyle 12}$} [c] at 3.00 -0.1
\put {${\scriptstyle 1}$} [c] at -0.15 1.00 
\endpicture$$
{\ninepoint  \baselineskip=10pt
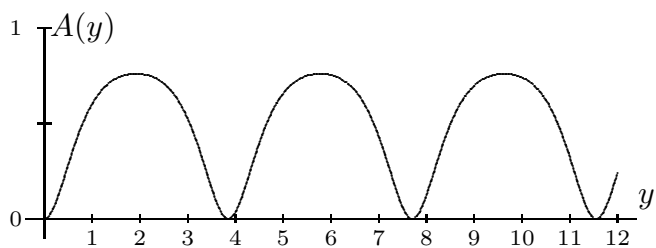
{\bf Figure 3.\/}  A periodic warp function $A(y)$ for $\Lambda=-1$,
$\lambda=-2$ and $\mu=-0.9$.  The initial condition is $A^{\prime\prime}(0) =
3.199870015$.  The value of the AdS length is $\ell=10$.}
\medskip

\listrefs
\bye